# A Causal Inference Approach to Eliminate the Impacts of Interfering Factors on Traffic Performance Evaluation

**Xiaobo Ma, Ph.D.**
Department of Civil & Architectural Engineering & Mechanics
The University of Arizona
1209 E 2nd St, Tucson, AZ 85721
Email: xiaoboma@arizona.edu

**Abolfazl Karimpour, Ph.D.**
College of Engineering
State University of New York Polytechnic Institute
100 Seymour Rd, Utica, NY 13502
Email: karimpa@sunypoly.edu

**Yao-Jan Wu, Ph.D., P.E.**
Department of Civil & Architectural Engineering & Mechanics
The University of Arizona
1209 E 2nd St, Tucson, AZ 85721
Email: yaojan@arizona.edu

**ABSTRACT**
Before and after study frameworks are widely adopted to evaluate the effectiveness of transportation policies and emerging technologies. However, many factors such as seasonal factors, holidays, and lane closure might interfere with the evaluation process by inducing variation in traffic volume during the before and after periods. In practice, limited effort has been made to eliminate the effects of these factors. In this study, an extreme gradient boosting (XGBoost)-based propensity score matching method is proposed to reduce the biases caused by traffic volume variation during the before and after periods. In order to evaluate the effectiveness of the proposed method, a corridor in the City of Chandler, Arizona where an advanced traffic signal control system has been recently implemented was selected. The results indicated that the proposed method is able to effectively eliminate the variation in traffic volume caused by the COVID-19 global Pandemic during the evaluation process. In addition, the results of the t-test and Kolmogorov-Smirnov (KS) test demonstrated that the proposed method outperforms other conventional propensity score matching methods. The application of the proposed method is also transferrable to other before and after evaluation studies and can significantly assist the transportation engineers to eliminate the impacts of traffic volume variation on the evaluation process.
**Keywords:** Before and after study, Propensity score matching, COVID-19

# 1. INTRODUCTION

The rapid expansion of urbanization, economic growth, and the increase in car ownership per capita, has significantly increased traffic congestion around the world (Wang et al., 2021). Therefore, transportation agencies and authorities are actively examining the potential benefit of implementing emerging technologies to tackle urban traffic congestion(Meng et al., 2016; Pan and Ryan, 2023a). Generally, in order to determine the benefits and evaluate the effectiveness of any technologies, experimental and non-experimental study designs are required to be developed(Pan and Ryan, 2023b).

Experimental designs are those in which the environment and variables are carefully controlled (Kim et al., 2020). Experimental designs are commonly used in medicine and social science fields and are intended to determine the cause and effect. The challenge in the experimental design procedure is that it is not always plausible to control all the variables involved in the environment and is less accurate for real-life setting scenarios. However, the most significant benefit of using an experimental design is that it provides specific conclusions (Li et al., 2018). Non-experimental designs are mainly taking place in a real-life setting where controlling all the factors in the test environment is not always feasible. One of the most common types of non-experimental design approaches is before and after studies. Before and after studies are generally utilized to evaluate the performance of introducing an intervention. Compared to other types of non-experimental designs, such as after-only and after-only-with-a-nonrandomized-control-group, the before and after study design provides more useful information about the effectiveness of the intervention (Robson, 2001).

The before and after study frameworks are comparably more straightforward, this approach has been commonly used in different real-life settings, including traffic and transportation evaluation projects(Luo et al., 2022; Ma, 2022; Ma et al., 2020a), such as ramp metering evaluation(Ma et al., 2020b), traffic safety study(Raihan et al., 2019), and traffic control device evaluation (Karimpour et al., 2021). Similar to other settings, while conducting a before and after study to evaluate the effectiveness of new transportation policies or emerging technologies many factors such as seasonal factors, holidays, lane closure, and work zone might interfere with the evaluation process by inducing variation in traffic volume. Without proactively resolving the potential biases caused by traffic volume variation, it is impossible to infer whether the observed differences in roadway traffic conditions are the true treatment effects. In practice, little effort has been made to eliminate the effects of these factors that might cause traffic volume variation. For instance, Lu et al. (2017) evaluated the Sydney Coordinated Adaptive Traffic System using taxi GPS data without considering the effects of traffic volume change during the study period (Lu et al., 2017). In another study, Remias et al. (2018) investigated the performance of coordinated signal timing using four types of data. However, the impact of traffic volume variation during before and after scenarios was not incorporated into the analysis (Remias et al., 2018). Bhouri et al. (2013) conducted field evaluations to explore the performances between isolated and coordinated ramp meterings without comparing traffic volume variation in volumes in before and after periods (Bhouri et al., 2013).

Furthermore, in many cases traffic volume reduction and traffic condition improvement are both observed in the after scenarios, with the traffic volume reduction, it is challenging to determine whether the traffic condition improvement is due to traffic volume variation or the impact of the implemented policy/technology. Similarly, in a scenario where the traffic volume increase and traffic condition deterioration are both observed in the after scenarios, it is challenging to determine whether the traffic condition deterioration is caused because of the traffic volume variation or the impact of the implemented policy/technology.

In before and after study frameworks, the assignment of sample is non-random across the treatment and control groups. Therefore, when estimating the effect of a treatment on an outcome, the result might be biased. One potential solution to resolve the biased caused by non-random assignment of treatment and control groups is using Propensity Score Matching (PSM). PSM Method has been applied in various fields, including transportation studies. For instance, Cao et al. (2010) used the PSM method to investigate how locations of people's residences in the local area affected the vehicle miles traveled (Cao et al., 2010). In another study, the PSM was adopted to examine in the regions that are in proximity to the bus stops, how much the improvements of bus stops were related to the transitions in ridership (Kim et al., 2020). To control for factors like socio-demographics, PSM has been employed in a study to investigate the impact of bus rapid transit (BRT) on walking and cycling (Chang et al., 2017). PSM has also been applied to examine China's massive high-speed rail (HSR) expansion and the added benefits evaluation through a lens of HSR-LCC (low-cost carriers) interactions (Wang et al., 2017). PSM was used to support a data-driven study about the impact of the "stay-at-home" policy on some specific modes of transportation in Manhattan, NYC (Lei and Ozbay, 2021). Applications of PSM have been seen in studies that explored the effects of the congestion charging plan in London (Ding et al., 2021) and rail services supported by the state in California (Talebian et al., 2018).

When using PSM methods, probit models are commonly used to estimate propensity scores. With only main effect terms, probit PSM methods generally provide adequate covariate balance. However, when covariates are highly correlated with each other, the bias-reducing capabilities of probit PSM methods substantially degrade. In contrast, regardless of sample size or the extent of non-additivity or non-linearity, machine learning-based PSM methods provided excellent performance in terms of covariate balance and effect estimation. In recent years, machine learning algorithms have been extensively applied to address a wide range of problems(Rubaiyat et al., 2018; Smith et al., 2023). The recent advancements in Artificial Intelligence (AI) and Machine Learning (ML) have empowered both researchers and practitioners to harness various machine learning techniques(Bao et al., 2022; Y. Yang et al., 2023). These techniques include random forest(Qu and Hickey, 2022), support vector machine(Wang and Qu, 2022), artificial neural network(He et al., 2020; Kong et al., 2022; Koome Murungi et al., 2023; Xu et al., 2023; Y.-X. Zhang et al., 2022), convolutional neural network(Liu et al., 2020; Tian et al., 2020; X. Yang et al., 2023; Yi and Qu, 2022; Zhang et al., 2023; Zhao and Melkote, 2022), generative adversarial network(Huang et al., 2020; Huang and Chiu, 2020), reinforcement learning(Chen et al., 2022; Fan et al., 2022; H. Zhou et al., 2023), curriculum learning(Dou et al., 2023), matric learning(Y. Zhang et al., 2022), representation learning(Dou et al., 2022a, 2022b), transformer(F. Zhou et al., 2023), and physics-informed neural networks(Huang and Wang, 2022). Moreover, the use of machine learning techniques to tackle intricate challenges in the transportation domain has experienced a surge in popularity in recent years(Zhang and Lin, 2022). Though machine learning methods have been widely used in transportation domain, machine learning-based PSM methods have been rarely applied to solve transportation related problems.

Although several studies explored the potential benefit of implementing new transportation policies or emerging technologies using PSM methods, very few studies focused on eliminating the biases originating from traffic volume variation on the evaluation results. In addition, determining a suitable sample size for the control group in PSM model development has not been studied before. Furthermore, a clear methodology to select the most accurate type of PSM method for transportation policy evaluation has not been studied previously. This study aims to fill these gaps by proposing a PSM-based framework that uses extreme gradient boosting (XGBoost) to estimate propensity scores. The proposed methodology applies t-test and Kolmogorov–Smirnov (KS) test to conduct a balancing test for determining the sample size for control groups. Besides, statistical analysis results from the t-test and KS test are employed to choose the most accurate one when comparing different PSM methods. The application of the proposed framework is then applied to a corridor in the City of Chandler, Arizona where an advanced traffic signal controller has been recently implemented.

## 2. METHODOLOGY

This study aims to reduce the biases caused by traffic volume variation during a before and after study evaluation using an XGBoost-based PSM method. If traffic volume variation exists between control and treatment groups, comparing the means of the outcomes between control and treatment groups will be misleading and the evaluation results will be biased. To eliminate the influence of traffic volume variation, an ''identical'' traffic volume condition in the control group should be identified in the treatment group. This matching is approximately equivalent to the process in which one of the two ''same'' traffic volume conditions is assigned into a treatment group and the other is assigned into a control group. By repeating this process for different traffic volume conditions in the treatment group, traffic volume conditions in the treatment group should not differ from those in the matched control group. That is, the matching process resembles an experiment with random assignment of treatment. Then, the effect of an intervention could be determined as the difference in mean outcomes between the matched control and treatment groups. In summary, the PSM method aims to create a counterfactual group that is statistically similar to the treatment group from the control group. Propensity scores are estimated using the XGBoost algorithm. The detailed explanations regarding PSM and XGBoost are shown in the following subsections.

### 2.1. Propensity Score Matching (PSM)
*2.1.1. Notations*

In causal inference, treatment evaluation is the estimation of the average effects of an intervention on the outcome of interest. It compares the outcomes between treated observations after the intervention and the control observations. The first step in the PSM method is assigning the observations into two groups: the treated group (receives the treatment) and the control group. In the PSM method, $D$ *is* the dependent variable and $x$ is the independent variable. $D$ is a binary variable that determines the type of observation. $D = 1$ means that the observation belongs to the treated group and $D = 0$ means that the observation belongs to the control group. Each observation has the conditional probability of receiving treatment given covariate $x$ as in Eq (1):

$$p(x) = prob(D = 1|x) = E(D|x) \tag{1}$$

Observations from treated and control groups are matched based on their propensity scores. Eq (2) denotes the outcomes ($y$) of the treated and control groups:

$$y = \begin{cases} y_1 \text{ if } D = 1 \\ y_0 \text{ if } D = 0 \end{cases} \tag{2}$$

In practice, the parameter of interest is usually the average treatment effect on the treated (ATET). ATET is the difference between the outcomes of the treated observations and the outcomes of the treated observations if they had not been treated, and can be calculated using Eq (3).

$$\text{ATET} = E(\Delta|D = 1) = E(y_1|x, D = 1) - E(y_0|x, D = 1) \tag{3}$$

The second term in Eq (3), is counterfactual, so it is unknown and needs to be estimated. After matching propensity scores, we can compare the outcomes of treated and matched control observations.

$$\text{ATET} = E(\Delta|p(x) = 1) = E(y_1|p(x), D = 1) - E(y_0|p(x), D = 1) \tag{4}$$

For the case of this study, ATET is the speed change and has the unit of mph.

*2.1.2. Assumptions*

To guarantee the results estimated by PSM are valid and accurate, there are three key assumptions:

1- Stable Unit Treatment Value Assumption (SUTVA)
   SUTVA requires that no intervention is imposed for any observations other than the treatment observations. SUTVA stands when: 1) Potential outcomes for any observations do not vary with the treatment assigned to other observations (no interference) and 2) For each observation, there are no different versions of each treatment level (no hidden variation of treatments).

2- Conditional Independence Assumption (CIA)
   The outcomes are independent of treatment, conditional on $x$. Meaning that the potential outcomes are independent of the treatment status after controlling for covariates $x$.
   
   $$y_0, y_1 \perp D|x \tag{5}$$

3- Common Support Condition (CSC)
   For each treated observation, there is a matched control observation with similar $x$. This is also known as overlap assumption which ensures there is enough overlap of observations between treated and control groups for accurate matching.
   
   $$0 < prob(D = 1|x) < 1 \tag{6}$$

*2.1.3. Propensity Score Estimation*

The propensity score is the likelihood that an individual observation receives the treatment. In this study, in order to estimate the propensity score, XGBoost is utilized. XGBoost improves the traditional gradient boosting decision tree algorithm by enhancing the computing time, generalization performance, and scalability (Mohammadi et al., 2019). The output of the XGBoost tree ensemble model with $K$ trees for each instance $i$ in the dataset is as below:

$$\hat{y}_i = \sum_{k=1}^{K} f_k(x_i) \tag{7}$$

Given a dataset with $n$ samples. Let $\hat{y}_i^{(t)}$ be the prediction of the $i$-th instance at the $t$-th iteration, we will need to add $f_t$ to minimize the following objective

$$\mathcal{L}^{(t)} = \sum_{i=1}^{n} l\left(y_i, \hat{y}_i^{(t-1)} + f_t(x_i)\right) + \Omega(f_t) \tag{8}$$

To quickly optimize the objective in the general setting, we approximate it using the second-order Taylor expansion.

$$\mathcal{L}^{(t)} \simeq \sum_{i=1}^{n} [l\left(y_i, \hat{y}_i^{(t-1)}\right) + g_i f_t(x_i) + \frac{1}{2} h_i f_t^2(x_i)] + \Omega(f_t) \tag{9}$$

where $g_i$ and $h_i$ are first and second-order gradient statistics on the loss function. We can remove the constant terms to obtain the following simplified objective at step $t$.

$$\tilde{\mathcal{L}}^{(t)} = \sum_{i=1}^{n} [g_i f_t(x_i) + \frac{1}{2} h_i f_t^2(x_i)] + \Omega(f_t) \tag{10}$$

let $I_j = \{i | q(x_i) = j\}$ be the instance set of leaf $j$. Eq (10) can be rewritten by expanding the regularization term $\Omega$ as follows

$$\tilde{\mathcal{L}}^{(t)} = \sum_{i=1}^{n} \left[g_i f_t(x_i) + \frac{1}{2} h_i f_t^2(x_i)\right] + \gamma T + \frac{1}{2} \lambda \sum_{j=1}^{T} \omega_j^2 \tag{11}$$

$$\tilde{\mathcal{L}}^{(t)} = \sum_{j=1}^{T} [\left(\sum_{i \in I_j} g_i\right) \omega_j + \frac{1}{2} (\sum_{i \in I_j} h_i + \lambda) \omega_j^2] + \gamma T \tag{12}$$

For a fixed structure $q(x)$, the optimal weight $\omega_j^*$ of leaf $j$ and the corresponding optimal objective function ($\tilde{\mathcal{L}}^{(t)}(q)$) can be computed as below:

$$\omega_j^* = -\frac{\sum_{i \in I_j} g_i}{\sum_{i \in I_j} h_i + \lambda} \tag{13}$$

$$\tilde{\mathcal{L}}^{(t)}(q) = -\frac{1}{2} \sum_{j=1}^{T} \frac{\left(\sum_{i \in I_j} g_i\right)^2}{\sum_{i \in I_j} h_i + \lambda} + \gamma T \tag{14}$$

Eq (14) can be used as a scoring function to measure the quality of a tree structure $q$. Usually, to enumerate all the tree structures $q$ of all the possibilities is of no chance. One possible greedy algorithm is employed alternatively. That is, to begin from one leaf and then iteratively increase branches to the whole tree. Assume that $I_L$ and $I_R$ are the instance sets of left and right nodes after the split. Letting $I = I_L \cup I_R$, then the loss reduction after the split is given by

$$\mathcal{L}_{split} = \frac{1}{2} \left[ \frac{\left(\sum_{i \in I_L} g_i\right)^2}{\sum_{i \in I_L} h_i + \lambda} + \frac{\left(\sum_{i \in I_R} g_i\right)^2}{\sum_{i \in I_R} h_i + \lambda} - \frac{(\sum_{i \in I} g_i)^2}{\sum_{i \in I} h_i + \lambda} \right] - \gamma \tag{15}$$

This formula is usually used in practice for evaluating the split candidates.

## 2.2. T-test and KS Test

In this study, the XGBoost-based propensity score matching (PSM) method was compared with two conventional propensity score matching methods: 1) PSM method with propensity scores estimated by probit regression (Probit), and 2) PSM method with propensity scores estimated by generalized boosted modeling (GBM). T-test and Kolmogorov-Smirnov Test were selected as the performance metrics. The T-test is commonly utilized to statistically compare the means between two groups of data(Khattak et al., 2020). The KS test is a non-parametric test that compares the similarity of two distributions (Hernandez and Hyun, 2020). The KS test calculates a distance between the Cumulative Distribution Functions (CDF) of two functions. KS statistic can be formulated as Eq (16):

$$D_{p,q} = \sup_x |F_p(x) - F_q(x)| \tag{16}$$

where $F_p(x)$ and $F_q(x)$ denote the CDFs estimated from $p$ and $q$ data observations, respectively.

### 2.3. Research Framework

Let $C_m$ denotes first $m$-week worth of data from the control group, $c_n$ denotes the $n$-th covariate, $T$ denotes $T$-week worth of data from the treatment group, and $N$ denotes the total number of covariates (in this study covariates are through movement traffic counts collected from the intersections on the study corridor). The procedures for using the PSM to construct the control group and evaluate the effects of transportation policies can be formulated as the following steps:

**Step 1:** Initially, $m$ and $n$ were set to be 1 as the searching process starts from the first week and first covariate. $C_m$ and $T$ were collected from the control group and treatment group.

**Step 2:** Covariates were selected to be included in the XGBoost model to estimate the propensity score for both control and treatment groups.

**Step 3:** The distributions of propensity scores were compared between control and treatment groups to check if the CSC was satisfied. If the CSC was not satisfied, then $C_m$ included in the XGBoost model needed to be re-selected.

**Step 4:** Nearest-Neighbor algorithm was selected to construct a counterfactual group from the control group.

**Step 5:** After matching using the Nearest-Neighbor algorithm, a series of t-tests and KS tests were conducted to check the difference of observations between the counterfactual group and treatment group for each covariate. This step could be regarded as a balancing test which was used to test if the counterfactual and treatment groups were statistically similar. If significant differences were found for any covariate, the $C_m$ needed to be re-specified and the process was repeated from the beginning.

**Step 6:** The effects of transportation policies could be evaluated by taking differences in counterfactual and treatment groups.

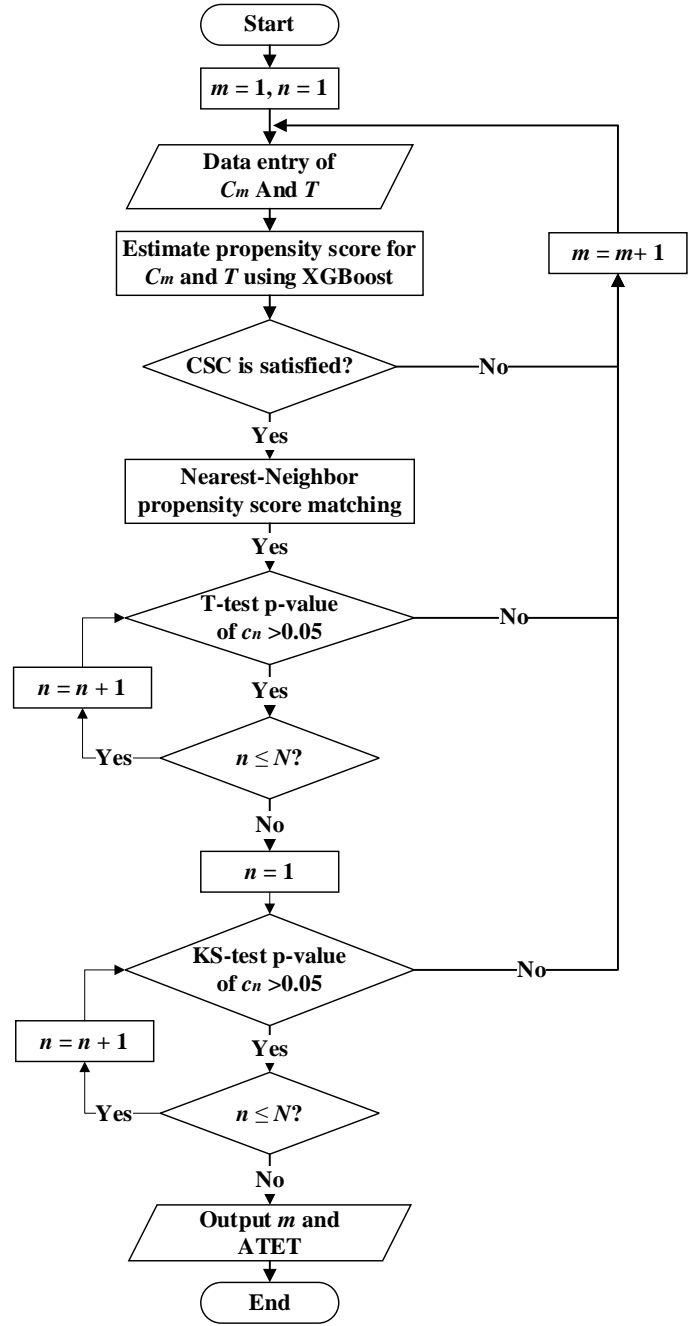
Figure 1 Research framework

## 3. CASE STUDY
In order to evaluate the effectiveness of the proposed method, a corridor in the City of Chandler, Arizona where an advanced traffic signal control system has been recently implemented was selected.

### 3.1. Study Sites
The Chandler Blvd. in the City of Chandler, Arizona has been selected as the study corridor. The traffic signal control on this corridor has recently been upgraded to a new traffic signal control system (the system was turned on during the week of Nov 9th, 2020). Chandler Blvd. is a west-east corridor connecting Interstate 10 to State Route 101 (Loop 101), with a speed limit of 45 mi/h. Since Chandler Fashion Center and Chandler Festival Shops are located along this corridor, Chandler Blvd. experiences high traffic demand during the year and especially during the holiday season. Figure 2 illustrates the study corridor and the intersections (numbered from 1 to 9) along this corridor that are equipped with this new traffic signal control system. City decision-makers decided to adopt this new technology to overcome the recurrent congestion happening at this corridor during holiday seasons. However, due to the COVID-19 global pandemic, the traffic volume and pattern changed significantly during the 2020 holiday season(M. Pan and Ryan, 2022; M. M. Pan and Ryan, 2022). The application of the proposed method will be used to determine whether this new traffic signal control system is able to effectively handle traffic demand and patterns during the 2020 holiday season.

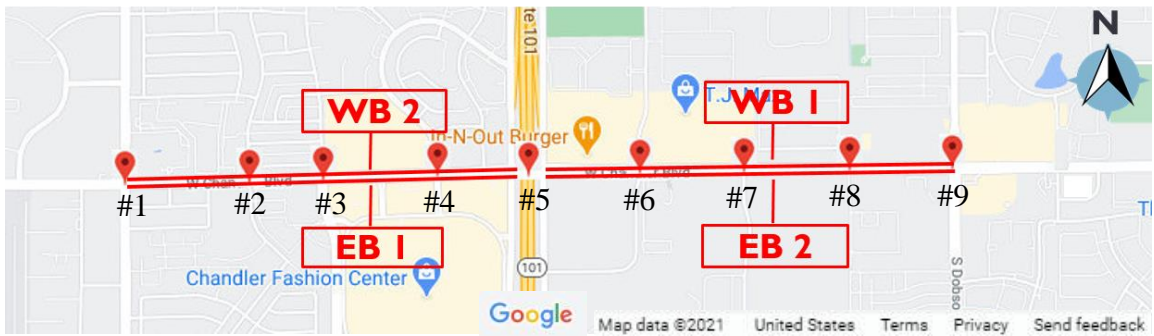

Figure 2 Layouts for Chandler Blvd (Image source: Google Maps)

### 3.2. Data Description

Two sources of data including sensor-based data and probe vehicle-based data were collected in this study.

1) Sensor-based data: The new traffic signal controller sensors are installed at each intersection approach and provide road user detection and classification via a fusion of machine vision and radar. The sensors are capable of classifying bikes, pedestrians, motorcycles, trucks, buses, cars, emergency vehicles, etc. In this study, only through movement traffic count of passenger cars were collected.
2) Probe vehicle-based data: The probe vehicle-based data used in this study is collected from INRIX data. INRIX data provides segment-based travel time and speed in a one-minute aggregation level at different Traffic Message Channel on the roadway. The data is collected from vehicles equipped with portable navigation systems or the smartphone inside the vehicle. For this study, speeds from four Traffic Message Channel segments were collected. WB1, WB2, EB1, and EB2 were used to differentiate these segments; these segments and their overlap with the study corridor are illustrated in Figure 2.

Sensor-based and probe vehicle-based data were collected from 6 am – 10 pm. To temporally match both data sources, turning movement count and speed data were aggregated into 1-hour time intervals.

### 3.3. Developing the Treatment and Control Groups

The first step in developing the PSM method is to build the treatment and control groups. The covariates used for the PSM method were through movement traffic counts at each intersection (EBT and WBT represent the eastbound and westbound through movement traffic counts of passenger cars, respectively). Through movement traffic counts from Christmas week of 2019 were selected and assigned to the control group (C1 in Table 1). For the treatment group, the turning movement counts after implementing the new traffic signal control system (Christmas week of 2020) were selected and assigned to the treatment group (T1 in Table 1).

Due to the COVID-19 global pandemic and its impact on the traffic volume and patterns, to make sure enough sample data is available for finding similar traffic conditions from the control group, an additional nine weeks of data from 2020 was collected and added to the control group (AC in Table 1).

Table 1 Observation duration for treatment and control groups

| Group | Week Number | Duration |
|---|---|---|
| | C1 | 2019-12-24~2019-12-30 |
| | AC1 | 2020-09-01~2020-09-07 |
| | AC2 | 2020-09-08~2020-09-14 |
| | AC3 | 2020-09-15~2020-09-21 |
| | AC4 | 2020-09-22~2020-09-28 |
| Control | AC5 | 2020-09-29~2020-10-05 |
| | AC6 | 2020-10-06~2020-10-12 |
| | AC7 | 2020-10-13~2020-10-19 |
| | AC8 | 2020-10-20~2020-10-26 |
| | AC9 | 2020-10-27~2020-11-02 |
| **Treatment** | T1 | 2020-12-20~2020-12-26 |

## 4. RESULTS
### 4.1. Propensity Score Matching Method Validation
*4.1.1. Validity of Common Support Condition (CSC)*

Before implementing the proposed method and evaluating the performance of the new traffic signal control system in the study corridor, the validity of the Common Support Condition (CSC) assumption needs to be investigated. CSC ensures there is enough overlap of observations between treated and control groups for good matching. The validity of the CSC can be checked by visually inspecting the spread of propensity scores for control and treatment groups. The boxplots illustrated in   show the propensity score distributions of control and treatment groups for each road segment.

In order for the CSC to be satisfied, the propensity score distributions (Figure 3) should have enough overlap for both control and treatment groups. Based on the boxplots provided in Figure 3, one can explore to what extent the propensity scores in control groups overlap with the propensity scores in treatment groups. Based on the box plot, it can be observed that the propensity scores from control and treatment groups have a significant overlap for all the four road segments, which indicates the CSC assumption is satisfied.

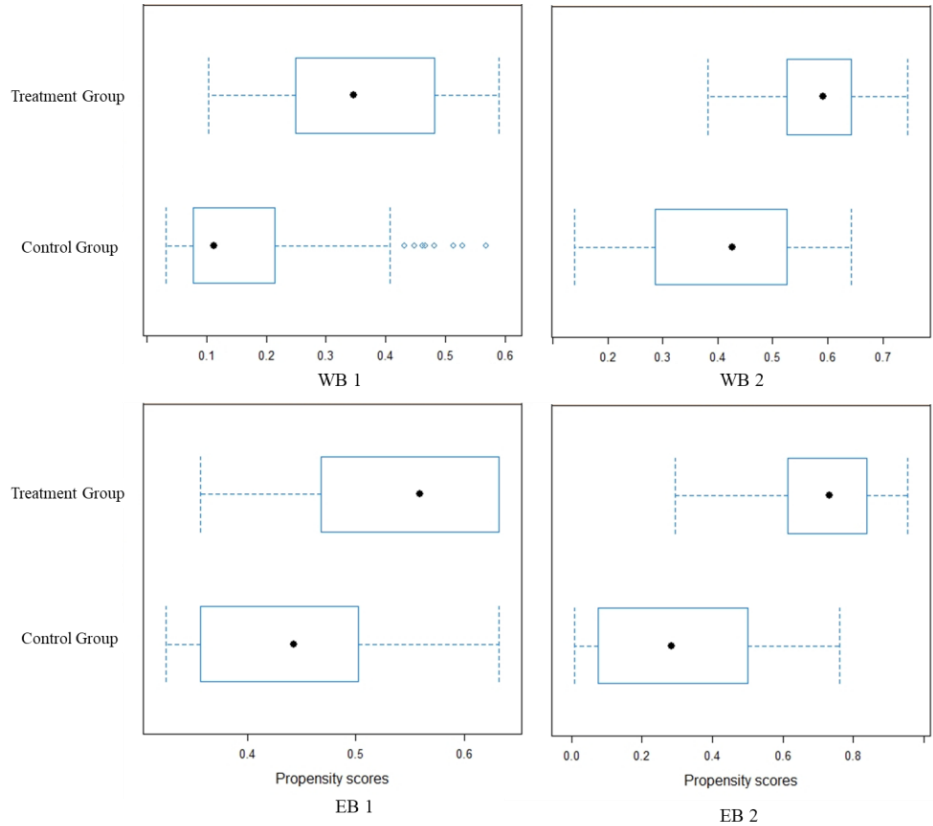

Figure 3 Propensity score distributions by treatment status

*4.2.2. Balancing Test*

The balancing test is another routine test that is used to check the validity of the propensity score matching method. As discussed in section 2.1.3, the PSM method aims to construct a counterfactual group that has no significant differences compared with the treatment group in terms of covariates' means (In this study, through movement traffic counts collected from study intersections). Traditionally, for the balancing test, t-test was used (Kim et al., 2020). T-test investigates whether there exists a statistically significant difference between the means of covariates' observations in the control and treatment groups, before and after matching. In this study, in addition to the t-test, the KS test was also incorporated to examine the distributions of the covariates' observations in the control and treatment groups.

The results of the t-test and KS test are provided in Table 2 and Table 3 for westbound and eastbound directions, respectively. Since the covariates selected for developing the PSM method were through movement counts, the results in these two tables are provided for each of the intersections on the study corridor. For instance, segment WB1 covers the westbound direction that includes intersections #5, #6, #7, #8, and #9. For instance, WBT #5 denote the westbound through movement traffic counts for intersection number 5 (refer to Figure 2). Then WBT#5, WBT#6, WBT#7, WBT#8, and WBT#9 were chosen as covariates in the PSM method. In these two tables, the null hypothesis of the KS test is that both samples in control and treatment groups come from a population with the same distribution. The null hypothesis of the t-test is that the means of two populations in control and treatment groups are equal.

Taking the first row of Table 2 as an example, for WBT#5 (westbound through movement traffic counts for intersection number 5), the difference of mean value between the treatment and control groups before matching was 32 veh/h (577-545), while after matching the difference reduced to 9 veh/h (577-568). That is, the proposed XGBoost-based PSM method significantly reduced the imbalance between treatment and control groups. Further, the KS test results show that after matching with the proposed XGBoost-based PSM method the CDF similarity increased between the treatment and control groups (KS statistic reduced from 0.21 to 0.08). The results of the t-test show that there is no significant difference (p>0.05) between covariate means of treatment and control groups before and after matching.

Results from Table 2 and Table 3 indicate that there is no significant difference between treatment and control groups in terms of covariate means and CDFs after matching. That is, all the covariates were balanced between treatment and counterfactual groups.

Table 2 Balancing test results of the treated and control groups for segments WB 1 and WB 2

| Measure | Segment WB 1 | | | Segment WB 2 | | |
|---|---|---|---|---|---|---|
| | Covariates | Before Matching | After Matching | Covariates | Before Matching | After Matching |
| Mean of treatment group | WBT#5 (veh/h) | 577 | 577 | WBT#1 (veh/h) | 599 | 599 |
| Mean of control group | | 545 | 568 | | 649 | 635 |
| T-test p-value | | 0.30 | 0.81 | | 0.17 | 0.34 |
| KS Statistic | | 0.21 | 0.08 | | 0.13 | 0.09 |
| KS test p-value | | <0.01 | 0.72 | | 0.35 | 0.78 |
| Mean of treatment group | WBT#6 (veh/h) | 798 | 798 | WBT#2 (veh/h) | 797 | 797 |
| Mean of control group | | 827 | 827 | | 769 | 769 |
| T-test p-value | | 0.48 | 0.55 | | 0.63 | 0.64 |
| KS Statistic | | 0.16 | 0.08 | | 0.13 | 0.12 |
| KS test p-value | | 0.02 | 0.69 | | 0.27 | 0.39 |
| Mean of treatment group | WBT#7 (veh/h) | 850 | 850 | WBT#3 (veh/h) | 730 | 730 |
| Mean of control group | | 841 | 845 | | 713 | 708 |
| T-test p-value | | 0.82 | 0.92 | | 0.67 | 0.60 |
| KS Statistic | | 0.18 | 0.07 | | 0.13 | 0.10 |
| KS test p-value | | <0.01 | 0.87 | | 0.35 | 0.67 |
| Mean of treatment group | WBT#8 (veh/h) | 906 | 906 | WBT#4 (veh/h) | 714 | 714 |
| Mean of control group | | 909 | 900 | | 734 | 725 |
| T-test p-value | | 0.94 | 0.91 | | 0.62 | 0.80 |
| KS Statistic | | 0.18 | 0.09 | | 0.10 | 0.08 |
| KS test p-value | | <0.01 | 0.55 | | 0.65 | 0.90 |
| Mean of treatment group | WBT#9 (veh/h) | 533 | 533 | WBT#5 (veh/h) | 577 | 577 |
| Mean of control group | | 473 | 531 | | 610 | 598 |
| T-test p-value | | 0.02 | 0.95 | | 0.44 | 0.63 |
| KS Statistic | | 0.29 | 0.09 | | 0.10 | 0.08 |
| KS test p-value | | <0.01 | 0.55 | | 0.65 | 0.88 |

Table 3 Balancing test results of the treated and control groups for segments EB 1 and EB 2

| Measure | Segment EB 1 | Segment EB 2 |
|---|---|---|

|  | Covariates | Before Matching | After Matching | Covariates | Before Matching | After Matching |
| --- | --- | --- | --- | --- | --- | --- |
| Mean of treatment group | EBT#1 (veh/h) | 597 | 597 | EBT#5 (veh/h) | 604 | 604 |
| Mean of control group |  | 660 | 614 |  | 611 | 588 |
| T-test p-value |  | 0.15 | 0.72 |  | 0.87 | 0.76 |
| KS Statistic |  | 0.16 | 0.06 |  | 0.12 | 0.09 |
| KS test p-value |  | 0.11 | 1.00 |  | 0.44 | 0.80 |
| Mean of treatment group | EBT#2 (veh/h) | 861 | 861 | EBT#6 (veh/h) | 738 | 738 |
| Mean of control group |  | 976 | 903 |  | 819 | 781 |
| T-test p-value |  | 0.07 | 0.56 |  | 0.09 | 0.47 |
| KS Statistic |  | 0.18 | 0.09 |  | 0.12 | 0.12 |
| KS test p-value |  | 0.06 | 0.80 |  | 0.44 | 0.42 |
| Mean of treatment group | EBT#3 (veh/h) | 745 | 745 | EBT#7 (veh/h) | 717 | 717 |
| Mean of control group |  | 731 | 690 |  | 748 | 732 |
| T-test p-value |  | 0.76 | 0.29 |  | 0.49 | 0.79 |
| KS Statistic |  | 0.13 | 0.11 |  | 0.07 | 0.10 |
| KS test p-value |  | 0.27 | 0.61 |  | 0.94 | 0.74 |
| Mean of treatment group | EBT#4 (veh/h) | 713 | 713 | EBT#8 (veh/h) | 907 | 907 |
| Mean of control group |  | 763 | 706 |  | 975 | 936 |
| T-test p-value |  | 0.28 | 0.90 |  | 0.26 | 0.70 |
| KS Statistic |  | 0.12 | 0.08 |  | 0.11 | 0.11 |
| KS test p-value |  | 0.44 | 0.91 |  | 0.54 | 0.52 |
| Mean of treatment group | EBT#5 (veh/h) | 604 | 604 | EBT#9 (veh/h) | 560 | 560 |
| Mean of control group |  | 611 | 563 |  | 500 | 506 |
| T-test p-value |  | 0.87 | 0.41 |  | 0.10 | 0.20 |
| KS Statistic |  | 0.12 | 0.09 |  | 0.15 | 0.12 |
| KS test p-value |  | 0.44 | 0.77 |  | 0.15 | 0.42 |

Based on the results from Figure 3, Tables 2 and 3, the CSC assumption and balancing test of the PSM have been validated. Next, in section 4.2, the application of the proposed method will be implemented on the study corridor.

## 4.2. Evaluating the Advanced Traffic Signal Control System

The proposed PSM method was used to determine the answer to the main question of our case study: Is this new traffic signal control system able to effectively handle traffic demand and patterns during the holiday season? For this study, the performance measure selected for evaluating the traffic signal control system is segment-based speed. After matching all the covariates were balanced between treatment and counterfactual groups. The effects of the new traffic signal control system could be evaluated by taking differences in counterfactual and treatment groups.

Determining the appropriate sample size for the control group plays an essential role in obtaining accurate evaluation results. Based on the research framework proposed in Figure 1, Table 4 presents the sample sizes needed for each road segment to evaluate the new traffic signal control system. Based on the sample size determination analysis, to create a counterfactual group that is statistically similar to the treatment group, five weeks of data is required for segment WB 1 and the remaining segments require only one week of data,-Identifying the correct sample size is significant to eliminate biases in the matching process.

Table 4 articulates the performance evaluation results for the road segments on the selected study corridor (refer to Figure 2). In this table, the effect denotes the change in segment-based speed after the new traffic signal control system was turned on (i.e., positive effects denote an increase in the segment-based speed). Based on the results, it can be concluded that the newly implemented traffic signal control system was able to improve the traffic conditions in the westbound direction by increasing the segment-based speed; the speed improvement was statistically significant at a 95% confidence level. However, for the eastbound direction, it was found that after implementing the new traffic signal control system, the segment-based speed was reduced. However, this reduction in speed was not statistically significant.

Table 4 New traffic signal control system performance evaluation results

| Road segment | Effect (mph) | T-test p-value | Sample size (week) |
|---|---|---|---|
| WB 1 | 1.7 | <0.01 | 5 |
| WB 2 | 1.5 | <0.01 | 1 |
| EB 1 | -1.0 | 0.14 | 1 |
| EB 2 | -0.3 | 0.58 | 1 |

### 4.3. Method Comparison

The functionality of the proposed method was compared with two conventional propensity score matching methods: 1) PSM method with propensity scores estimated by probit regression (Probit), and 2) PSM method with propensity scores estimated by generalized boosted modeling (GBM). The difference between the proposed method and these two conventional methods is in the algorithm they use for estimating the propensity scores.

Table 5 illustrates the comparison result of different propensity score matching methods. It can be observed that for two out of four road segments (WB 1 and EB 1), the probit-based PSM method is unable to construct counterfactual groups that are statistically similar to treatment groups. For the remaining segments (WB 2 and EB 2), the summary statistics of the probit-based PSM method are shown in Table 6 and Table 7.

While comparing the proposed method (XGBoost-based PSM method) with the GBM-based PSM method, both methods obtain similar results for three out of four segments (WB 1, WB 2, and EB 1). However, the results are different for segment EB2. To determine which method produced more reliable results, the summary statistics of GBM-based and XGBoost-based PSM methods for segment EB 2 are displayed in Table 8.

Table 5 Results comparison between different methods

| Methods | Measures | Segment WB 1 | Segment WB 2 | Segment EB 1 | Segment EB 2 |
|---|---|---|---|---|---|
| Probit | Effects (mph) | − | 0.8 | − | -0.6 |
| | T-test p-value | − | 0.25 | − | 0.32 |
| | Sample size (week) | − | 3 | − | 2 |
| GBM | Effects (mph) | 1.7 | 1.3 | -0.9 | -0.8 |
| | T-test p-value | <0.01 | 0.01 | 0.18 | 0.04 |
| | Sample size (week) | 5 | 1 | 1 | 1 |
| XGBoost | Effects (mph) | 1.7 | 1.5 | -1.0 | -0.3 |
| | T-test p-value | <0.01 | <0.01 | 0.14 | 0.58 |
| | Sample size (week) | 5 | 1 | 1 | 1 |

Table 6 and Table 7 summarize the effect of adding data (in terms of the week) on the t-test p-values, KS statistics, KS test p-values, and estimated effects using the probit-based PSM method while evaluating performance on segment WB 2 and segment EB 2. Based on the results from these two tables, it can be observed that three and two weeks of data are required for the probit-based PSM methods to construct counterfactual groups that are statistically similar to the treatment groups, for WB 2 and EB 2 directions, respectively.

It is worth mentioning that, if the t-test was the only statistical test that was utilized to conduct the balancing test, then only one week of data was needed for both segments (WB 2 and EB 2). However, in this study, the KS test was also utilized to further examine the difference between the distributions of the counterfactual groups and treatment groups. Incorporating the results from the KS test, it was found that in order to construct counterfactual groups that are statistically similar to the treatment groups, three weeks of data is required for WB 2 and two weeks of data is required for the EB 2. Comparing these results for WB 2 and EB 2 with the results from the proposed method and the GBM-based PSM method, we can conclude that the probit-based PSM methods require more data to construct counterfactual groups.

Table 6 Summary statistics of probit-based PSM method for segment WB 2

| Sample Size (Week) | Variable | After Matching | | | Effect (mph) | T-test p-value |
|---|---|---|---|---|---|---|
| | | T-test p-value | KS Statistic | KS test p-value | | |
| 1 | WBT #1 | 0.39 | 0.14 | 0.21 | 1.5 | 0.13 |
| | WBT #2 | 0.07 | 0.15 | 0.16 | | |
| | WBT #3 | 0.51 | 0.13 | 0.28 | | |
| | WBT #4 | 0.30 | 0.16 | 0.12 | | |
| | WBT #5 | 0.19 | 0.18 | 0.04 | | |
| 2 | WBT #1 | 0.36 | 0.10 | 0.59 | 1.5 | 0.06 |
| | WBT #2 | 0.11 | 0.15 | 0.13 | | |
| | WBT #3 | 0.61 | 0.08 | 0.80 | | |
| | WBT #4 | 0.56 | 0.08 | 0.80 | | |
| | WBT #5 | 0.27 | 0.19 | 0.02 | | |
| 3 | WBT #1 | 0.71 | 0.13 | 0.24 | 0.8 | 0.25 |
| | WBT #2 | 0.85 | 0.11 | 0.38 | | |
| | WBT #3 | 0.46 | 0.12 | 0.30 | | |
| | WBT #4 | 0.51 | 0.11 | 0.38 | | |
| | WBT #5 | 0.82 | 0.12 | 0.30 | | |

Table 7 Summary statistics of probit-based PSM method for segment EB 2

| Sample Size (Week) | Variable | After Matching | | | Effect (mph) | T-test p-value |
|---|---|---|---|---|---|---|
| | | T-test p-value | KS Statistic | KS test p-value | | |
| 1 | EBT #5 | 0.62 | 0.22 | 0.01 | -1.3 | 0.30 |
| | EBT #6 | 0.37 | 0.25 | <0.01 | | |
| | EBT #7 | 0.71 | 0.19 | 0.03 | | |
| | EBT #8 | 0.08 | 0.22 | 0.01 | | |
| | EBT #9 | 0.33 | 0.21 | 0.01 | | |
|   | EBT #5 | 0.55 | 0.13 | 0.31 |   |   |
|   | EBT #6 | 0.44 | 0.15 | 0.13 |   |   |
| 2 | EBT #7 | 0.52 | 0.17 | 0.07 | -0.6 | 0.32 |
|   | EBT #8 | 0.50 | 0.18 | 0.06 |   |   |
|   | EBT #9 | 0.46 | 0.16 | 0.10 |   |   |

Table 8 provides summary statistics of GBM-based and XGBoost-based PSM methods for segment EB 2. Mean values, t-test p-values, KS statistics, and KS test p-values between treatment groups and control groups before and after matching are compared. From the table, it can be seen that after matching, the XGBoost-based PSM method acquires the lower KS statistics and more accurate mean values of control groups in most cases after matching, which means the XGBoost-based PSM method outperforms the GBM-based PSM method regarding reducing the imbalance between treatment and control groups. In summary, compared with other methods, estimating propensity scores using XGBoost can effectively match treatment and control groups and accurately measure the ATET.

Table 8 Summary statistics of GBM and XGBoost methods for segment EB 2

| Variable | Measures | Before Matching | After Matching | |
|---|---|---|---|---|
|   |   |   | GBM | XGBoost |
| EBT #5 (veh/h) | Mean of treatment group | 604 | 604 | 604 |
|   | Mean of control group | 611 | **611** | 588 |
|   | T-test p-value | 0.87 | 0.87 | 0.76 |
|   | KS Statistic | 0.12 | 0.12 | **0.09** |
|   | KS test p-value | 0.44 | 0.44 | 0.80 |
| EBT #6 (veh/h) | Mean of treatment group | 738 | 738 | 738 |
|   | Mean of control group | 819 | 819 | **781** |
|   | T-test p-value | 0.09 | 0.09 | 0.47 |
|   | KS Statistic | 0.12 | 0.12 | **0.12** |
|   | KS test p-value | 0.44 | 0.44 | 0.42 |
| EBT #7 (veh/h) | Mean of treatment group | 717 | 717 | 717 |
|   | Mean of control group | 748 | 748 | **732** |
|   | T-test p-value | 0.49 | 0.49 | 0.79 |
|   | KS Statistic | 0.07 | **0.07** | 0.10 |
|   | KS test p-value | 0.94 | 0.94 | 0.74 |
| EBT #8 (veh/h) | Mean of treatment group | 907 | 907 | 907 |
|   | Mean of control group | 975 | 975 | **936** |
|   | T-test p-value | 0.26 | 0.26 | 0.70 |
|   | KS Statistic | 0.11 | 0.11 | **0.11** |
|   | KS test p-value | 0.54 | 0.54 | 0.52 |
| EBT #9 (veh/h) | Mean of treatment group | 560 | 560 | 560 |
|   | Mean of control group | 500 | 500 | **506** |
|   | T-test p-value | 0.10 | 0.10 | 0.20 |
|   | KS Statistic | 0.15 | 0.15 | **0.12** |
|   | KS test p-value | 0.15 | 0.15 | 0.42 |

# 5. CONCLUSION

Generally, to conduct performance evaluation, before and after study frameworks are developed. However, many factors such as seasonal factors, holidays, lane closure, and work zone might interfere with the evaluation process by inducing variation in traffic volume during before and after study periods. If traffic volume variation exists between control and treatment groups, comparing the means of the outcomes between control and treatment groups will be misleading and the evaluation results will be biased. In this study, an XGBoost-based PSM method was proposed to potentially reduce the biases caused by traffic volume variation during a before and after study evaluation.

The application of the proposed method was used to evaluate a newly implemented traffic signal control system on a corridor in the City of Chandler, Arizona. The results indicated that the proposed XGBoost-based PSM method is able to effectively eliminate the variation in traffic volume caused by the COVID-19 global Pandemic during the evaluation process. In addition, the results of the t-test and Kolmogorov-Smirnov (KS) test demonstrated that the proposed method outperforms other conventional propensity score matching methods. The application of the proposed method is also transferrable to other before and after evaluation studies and can significantly assist the transportation engineers to eliminate the impacts of traffic volume variation on the evaluation process.

As the proposed framework has high flexibility that allows the incorporation of multiple covariates, future work could consider adding more variables in the proposed XGBoost-based PSM method. Also, weather conditions, incidents, and other factors that might impact the traffic operation can be included in the proposed method to further improve the generalization ability of the method. This paper is the first attempt to use XGBoost to estimate propensity score and try to minify the performance evaluation errors caused by traffic volume variation using the PSM method. In the future, more advanced machine learning methods and more sophisticated input features could be applied to further improve the estimation performance.

**ACKNOWLEDGMENTS**

The authors would like to thank the Maricopa Association of Governments for funding support. We are also grateful for the data support from the Arizona Department of Transportation and the City of Chandler, Arizona.

# REFERENCES

Bao, R., Wu, X., Xian, W., Huang, H., 2022. Doubly sparse asynchronous learning for stochastic composite optimization, in: Proceedings of the Thirty-First International Joint Conference on Artificial Intelligence, IJCAI. pp. 1916–1922.

Bhouri, N., Haj-Salem, H., Kauppila, J., 2013. Isolated versus coordinated ramp metering: Field evaluation results of travel time reliability and traffic impact. Transp. Res. Part C Emerg. Technol. 28, 155–167. https://doi.org/10.1016/j.trc.2011.11.001

Cao, X. (Jason), Xu, Z., Fan, Y., 2010. Exploring the connections among residential location, self-selection, and driving: Propensity score matching with multiple treatments. Transp. Res. Part A Policy Pract. 44, 797–805. https://doi.org/10.1016/j.tra.2010.07.010

Chang, A., Miranda-Moreno, L., Cao, J., Welle, B., 2017. The effect of BRT implementation and streetscape redesign on physical activity: A case study of Mexico City. Transp. Res. Part A Policy Pract. 100, 337–347. https://doi.org/10.1016/j.tra.2017.04.032

Chen, Jiayu, Chen, Jingdi, Lan, T., Aggarwal, V., 2022. Scalable multi-agent covering option discovery based on kronecker graphs. Adv. Neural Inf. Process. Syst. 35, 30406–30418.

Ding, H., Sze, N.N., Li, H., Guo, Y., 2021. Affected area and residual period of London Congestion Charging scheme on road safety. Transp. Policy 100, 120–128. https://doi.org/10.1016/j.tranpol.2020.10.012

Dou, J.X., Bao, R., Song, S., Yang, S., Zhang, Y., Liang, P.P., Mao, H.H., 2023. Demystify the Gravity Well in the Optimization Landscape (student abstract), in: Proceedings of the AAAI Conference on Artificial Intelligence.

Dou, J.X., Jia, M., Zaslavsky, N., Ebeid, M., Bao, R., Zhang, S., Ni, K., Liang, P.P., Mao, H., Mao, Z.-H., 2022a. Learning more effective cell representations efficiently, in: NeurIPS 2022 Workshop on Learning Meaningful Representations of Life.

Dou, J.X., Pan, A.Q., Bao, R., Mao, H.H., Luo, L., 2022b. Sampling through the lens of sequential decision making. arXiv Prepr. arXiv2208.08056.

Fan, X., Zhang, X., Yu, X., 2022. A graph convolution network-deep reinforcement learning model for resilient water distribution network repair decisions. Comput. Civ. Infrastruct. Eng. 37, 1547–1565.

He, S., Pepin, L., Wang, G., Zhang, D., Miao, F., 2020. Data-driven distributionally robust electric vehicle balancing for mobility-on-demand systems under demand and supply uncertainties, in: 2020 IEEE/RSJ International Conference on Intelligent Robots and Systems (IROS). pp. 2165–2172.

Hernandez, S., Hyun, K., 2020. Fusion of weigh-in-motion and global positioning system data to estimate truck weight distributions at traffic count sites. J. Intell. Transp. Syst. Technol. Planning, Oper. 24, 201–215. https://doi.org/10.1080/15472450.2019.1659793

Huang, B., Wang, J., 2022. Applications of physics-informed neural networks in power systems-a review. IEEE Trans. Power Syst. 38, 572–588.

Huang, Z., Arian, A., Yuan, Y., Chiu, Y.C., 2020. Using Conditional Generative Adversarial Nets and Heat Maps with Simulation-Accelerated Training to Predict the Spatiotemporal Impacts of Highway Incidents. Transp. Res. Rec. 2674, 836–849. https://doi.org/10.1177/0361198120925069

Huang, Z.R., Chiu, Y.-C., 2020. Spatiotemporal Nonrecurring Traffic Spillback Pattern Prediction for Freeway Merging Bottleneck Using Conditional Generative Adversarial Nets with Simulation Accelerated Training, in: 2020 IEEE 23rd International Conference on Intelligent Transportation Systems (ITSC). pp. 1–6.

Karimpour, A., Kluger, R., Liu, C., Wu, Y.J., 2021. Effects of speed feedback signs and law enforcement on driver speed. Transp. Res. Part F Traffic Psychol. Behav. 77, 55–72. https://doi.org/10.1016/j.trf.2020.11.011

Khattak, Z.H., Magalotti, M.J., Fontaine, M.D., 2020. Operational performance evaluation of adaptive traffic control systems: A Bayesian modeling approach using real-world GPS and private sector PROBE data. J. Intell. Transp. Syst. Technol. Planning, Oper. 24, 156–170. https://doi.org/10.1080/15472450.2019.1614445

Kim, J.Y., Bartholomew, K., Ewing, R., 2020. Another one rides the bus? The connections between bus stop amenities, bus ridership, and ADA paratransit demand. Transp. Res. Part A Policy Pract. 135, 280–288. https://doi.org/10.1016/j.tra.2020.03.019

Kong, R., Zhang, C.C., Sun, R., Chhabra, V., Nadimpalli, T., Konstan, J.A., 2022. Multi-Objective Personalization in Multi-Stakeholder Organizational Bulk E-mail: A Field Experiment. Proc. ACM Human-Computer Interact. 6, 1–27.

Koome Murungi, N., Pham, M.V., Dai, X., Qu, X., 2023. Trends in Machine Learning and Electroencephalogram (EEG): A Review for Undergraduate Researchers. arXiv e-prints arXiv--2307.

Lei, Y., Ozbay, K., 2021. A robust analysis of the impacts of the stay-at-home policy on taxi and Citi Bike usage : A case study of Manhattan. Transp. Policy 110, 487–498. https://doi.org/10.1016/j.tranpol.2021.07.003

Li, H., Ding, H., Ren, G., Xu, C., 2018. Effects of the London Cycle Superhighways on the usage of the London Cycle Hire. Transp. Res. Part A Policy Pract. 111, 304–315. https://doi.org/10.1016/j.tra.2018.03.020

Liu, D., Cui, Y., Chen, Y., Zhang, J., Fan, B., 2020. Video object detection for autonomous driving: Motion-aid feature calibration. Neurocomputing 409, 1–11.

Lu, Y.C., Krambeck, H., Tang, L., 2017. Use of big data to evaluate and improve performance of traffic signal systems in resource-constrained countries evidence from Cebu city, Philippines. Transp. Res. Rec. 2620, 20–30. https://doi.org/10.3141/2620-03

Luo, X., Ma, X., Munden, M., Wu, Y.-J., Jiang, Y., 2022. A Multisource Data Approach for Estimating Vehicle Queue Length at Metered On-Ramps. J. Transp. Eng. Part A Syst. 148, 1–9. https://doi.org/10.1061/jtepbs.0000622

Ma, X., 2022. Traffic Performance Evaluation Using Statistical and Machine Learning Methods. The University of Arizona.

Ma, X., Karimpour, A., Wu, Y., 2020a. Statistical evaluation of data requirement for ramp metering performance assessment. Transp. Res. Part A 141, 248–261. https://doi.org/10.1016/j.tra.2020.09.011

Ma, X., Karimpour, A., Wu, Y., 2020b. Statistical evaluation of data requirement for ramp metering performance assessment. Transp. Res. Part A 141, 248–261. https://doi.org/10.1016/j.tra.2020.09.011

Meng, X., Tang, S., Liu, X., Zhang, L., 2016. Oversaturated traffic signal optimization based on active control, in: 2016 IEEE International Conference on Intelligent Transportation Engineering (ICITE). pp. 91–95.

Mohammadi, R., He, Q., Ghofrani, F., Pathak, A., Aref, A., 2019. Exploring the impact of foot-by-foot track geometry on the occurrence of rail defects. Transp. Res. Part C Emerg. Technol. 102, 153–172. https://doi.org/10.1016/j.trc.2019.03.004

Pan, M., Ryan, A., 2023a. Segmenting the target audience for transportation demand management programs: An investigation between mode shift and individual characteristics. Int. J. Sustain. Transp. 1–22.

Pan, M., Ryan, A., 2023b. How to select distracted driving countermeasures evaluation metrics: A systematic review. J. Transp. Saf. \& Secur. 1–31.

Pan, M., Ryan, A., 2022. The Impact of Confirmation Bias on Perceived Health Risk of Using Public Transit: An Evaluation during the Pandemic. J. Transp. \& Heal. 25, 101428.

Pan, M.M., Ryan, A., 2022. Investigating confirmation bias in transportation: An analysis of perceived health risk on public transit during the pandemic. J. Transp. \& Heal. 26, 101485.

Qu, X., Hickey, T.J., 2022. Eeg4home: A human-in-the-loop machine learning model for eeg-based bci, in: International Conference on Human-Computer Interaction. pp. 162–172.

Raihan, M.A., Alluri, P., Wu, W., Gan, A., 2019. Estimation of bicycle crash modification factors (CMFs) on urban facilities using zero inflated negative binomial models. Accid. Anal. Prev. 123, 303–313. https://doi.org/10.1016/j.aap.2018.12.009

Remias, S.M., Day, C.M., Waddel, J.M., Kirsch, J.N., Trepanier, T., 2018. Evaluating the performance of coordinated signal timing: Comparison of common data types with automated vehicle location data. Transp. Res. Rec. 2672, 128–132. https://doi.org/10.1177/0361198118794546

Robson, L.S., 2001. Guide to evaluating the effectiveness of strategies for preventing work injuries; how to show whether a safety invervention really works.

Rubaiyat, A.H.M., Qin, Y., Alemzadeh, H., 2018. Experimental resilience assessment of an open-source driving agent, in: 2018 IEEE 23rd Pacific Rim International Symposium on Dependable Computing (PRDC). pp. 54–63.

Smith, W., Qin, Y., Singh, S., Burke, H., Furukawa, T., Dissanayake, G., 2023. A Multistage Framework for Autonomous Robotic Mapping with Targeted Metrics. Robotics 12, 39.

Talebian, A., Zou, B., Hansen, M., 2018. Assessing the impacts of state-supported rail services on local population and employment: A California case study. Transp. Policy 63, 108–121. https://doi.org/10.1016/j.tranpol.2017.12.013

Tian, Y., Deng, X., Zhu, Y., Newsam, S., 2020. Cross-time and orientation-invariant overhead image geolocalization using deep local features, in: Proceedings of the IEEE/CVF Winter Conference on Applications of Computer Vision. pp. 2512–2520.

Wang, K., Xia, W., Zhang, A., 2017. Should China further expand its high-speed network? Consider the low-cost carrier factor. Transp. Res. Part A Policy Pract. 100, 105–120. https://doi.org/10.1016/j.tra.2017.04.010

Wang, R., Qu, X., 2022. EEG daydreaming, a machine learning approach to detect daydreaming activities, in: International Conference on Human-Computer Interaction. pp. 202–212.

Wang, X., Rodríguez, D.A., Mahendra, A., 2021. Support for market-based and command-and-control congestion relief policies in Latin American cities: Effects of mobility, environmental health, and city-level factors. Transp. Res. Part A Policy Pract. 146, 91–108. https://doi.org/10.1016/j.tra.2020.12.004

Xu, N., Nie, Q., Liu, J., Jones, S., 2023. Post-pandemic shared mobility and active travel in Alabama: A machine learning analysis of COVID-19 survey data. Travel Behav. Soc. 32, 100584.

Yang, X., Bist, R., Subedi, S., Wu, Z., Liu, T., Chai, L., 2023. An automatic classifier for monitoring applied behaviors of cage-free laying hens with deep learning. Eng. Appl. Artif. Intell. 123, 106377.

Yang, Y., Liu, C., Zhang, Z., 2023. Particle-based online bayesian sampling. arXiv Prepr. arXiv2302.14796.

Yi, L., Qu, X., 2022. Attention-Based CNN Capturing EEG Recording's Average Voltage and Local Change, in: International Conference on Human-Computer Interaction. pp. 448–459.

Zhang, D., Zhou, F., Jiang, Y., Fu, Z., 2023. Mm-bsn: Self-supervised image denoising for real-world with multi-mask based on blind-spot network, in: Proceedings of the IEEE/CVF Conference on Computer Vision and Pattern Recognition. pp. 4188–4197.

Zhang, L., Lin, W.-H., 2022. Calibration-free Traffic Signal Control Method Using Machine Learning Approaches, in: 2022 International Conference on Electrical, Computer and Energy Technologies (ICECET). pp. 1–6.

Zhang, Y.-X., Chen, F.-Y., Feng, Q.-K., Liu, D.-F., Pei, J.-Y., Zhong, S.-L., Yang, Z., Dang, Z.-M., 2022. Artificial intelligence aided design for film capacitors, in: 2022 IEEE International Conference on High Voltage Engineering and Applications (ICHVE). pp. 1–4.

Zhang, Y., Bao, R., Pei, J., Huang, H., 2022. Toward Unified Data and Algorithm Fairness via Adversarial Data Augmentation and Adaptive Model Fine-tuning, in: 2022 IEEE International Conference on Data Mining (ICDM). pp. 1317–1322.

Zhao, C., Melkote, S.N., 2022. Learning the Part Shape and Part Quality Generation Capabilities of Machining and Finishing Processes Using a Neural Network Model, in: International Design Engineering Technical Conferences and Computers and Information in Engineering Conference. p. V002T02A010.

Zhou, F., Zhang, D., Fu, Z., 2023. High dynamic range imaging with context-aware transformer. arXiv Prepr. arXiv2304.04416.

Zhou, H., Lan, T., Aggarwal, V., 2023. Value functions factorization with latent state information sharing in decentralized multi-agent policy gradients. IEEE Trans. Emerg. Top. Comput. Intell.